\documentclass[aip,reprint]{revtex4-1}

\draft 
\usepackage{graphicx}

\begin{document}


\title{Heterogeneous nucleation and heat flux avalanches in La(Fe,Si)$_{13}$ magnetocaloric compounds near the critical point. } 



\author{C.Bennati}
\affiliation{Istituto Nazionale di Ricerca Metrologica (INRIM), Strada delle Cacce 91, 10135 Turin, Italy}
\affiliation{Department of Applied Science and Technology, Politecnico di Torino, C.so Duca degli Abruzzi 24, 10129 Turin, Italy}
\author{L.Gozzelino}
\affiliation{Department of Applied Science and Technology, Politecnico di Torino, C.so Duca degli Abruzzi 24, 10129 Turin, Italy}
\author{E.S.Olivetti}
\affiliation{Istituto Nazionale di Ricerca Metrologica (INRIM), Strada delle Cacce 91, 10135 Turin, Italy}
\author{V.Basso}
\affiliation{Istituto Nazionale di Ricerca Metrologica (INRIM), Strada delle Cacce 91, 10135 Turin, Italy}

\date{\today}

\begin{abstract}

The phase transformation kinetics of LaFe$_{11.41}$Mn$_{0.30}$Si$_{1.29}$-H$_{1.65}$ magnetocaloric compound is addressed by low rate calorimetry experiments. Scans at 1 mK/s show that its first order phase transitions are made by multiple heat flux avalanches. Getting very close to the critical point, the step-like discontinuous behaviour associated with avalanches is smoothed out and thermal hysteresis disappears. This result is confirmed by magneto-resistivity measurements and allows to measure accurate values of the zero field hysteresis ($\Delta T_{hyst}$ = 0.37 K) and of the critical field ($H_c$ = 1.19 T). The number and magnitude of heat flux avalanches change with magnetic field, showing the interplay between the intrinsic energy barrier between phases and the microstructural disorder of the sample.
\end{abstract}


\maketitle 

A strong attention is nowadays directed to room temperature refrigeration techniques based on the magnetocaloric effect (MCE) because they allow a reduced energy consumption and a lower environmental impact with respect to gas compression technologies \cite*{liu2012giant}$^,$\cite*{tegus2002transition}.
A class of materials which are promising candidates for magnetic cooling, is the one based on the La(Fe,Si)$_{13}$ compound \cite*{fujieda2002large}. These intermetallics show a large MCE because they exploit a sharp drop in magnetization associated with a ferromagnetic (FM) to paramagnetic (PM) phase transition. Near the transition temperature, $T_t$, magnetic fields of about 2 T can provide an adiabatic temperature variation up to $\Delta T_{ad}$ = 7 K \cite*{fujieda2002large}. This giant MCE, observed in magnetic transitions of the first order type, implies thermo-magnetic hysteresis as a drawback for applications. Understanding the mechanism which underlies thermo-magnetic hysteresis is thus of great importance for the modelling of magnetic refrigeration cycles. 
It is known that the hysteresis width of La(Fe,Si)$_{13}$ is influenced by the strength of the magnetic field \cite*{fujita2001itinerant}, by hydrostatic pressure \cite*{fujita2012stability} and by substitution element at Fe sites. Moreover, as pointed out on several works\cite*{morrison2009capturing}$^,$\cite*{niemann2016reducing}, and particularly on those regarding La(Fe,Si)$_{13}$ based materials \cite*{kuepferling2014dynamics}, the transformation process of first order magnetocaloric materials is due to the motion of phase boundaries between FM and PM phases. This motion takes place on a complex energy landscape influenced by several factors. For example, the strains generated by the lattice shrinking at the PM/FM transitions may influence the free energy profile at local site \cite*{waske2015asymmetric} as well as the magnetic and structural disorder which can block or favour the transition front advance\cite*{bennati2015local}$^,$\cite*{kim2011magnetocaloric} .

We address this issue by investigating the in-temperature transformation process of a LaFe$_{11.41}$Mn$_{0.30}$Si$_{1.29}$-H$_{1.65}$ sample.
The chosen composition has a transition on the border between first and second order types and it represents the best compromise for application near room temperature due to the large $\Delta S_{iso}$ (19 J kg$^{-1}$ K$^{-1}$) and the low zero field thermal hysteresis (0.4 K at $T_t \approx$ 295 K) \cite*{basso2015specific}.
By exploiting low (1 mK/s) and fast (up to 100 mK/s) temperature rates calorimetry experiments, and by using electrical resistivity measurements, we are able to show that the phase transformation is associated to an heterogeneous nucleation/pinning mechanism characterized by a repeatable sequence of heat flux avalanches\cite*{kuepferling2014dynamics}. When a constant external magnetic field is applied and the transition is shifted to higher temperatures, the avalanches change in number and decrease in amplitude until they finally disappear above the critical point. 
On the basis of the magnetic phase diagram reported in \cite*{basso2015specific}, the zero field $\Delta T_{hyst} = 0.4$ K of the nominal composition LaFe$_{11.41}$Mn$_{0.30}$Si$_{1.29}$-H$_{1.65}$ can be suppressed with a magnetic field of 1.3 T. 

\medskip

The starting composition is produced by powder metallurgy by Vacuumschmelze GmbH $\&$ Co\cite*{barcza2011stability}: powders of the ternary La-Fe-Si alloy are blended with Mn-rich powders. The blends are compacted by cold isostatic pressing and sintered at 1353 K, obtaining fully dense materials with density of about 7.2 g/cm$^3$, characterized by La(Fe,Si)$_{13}$ grains of several tens of micrometers and a minor amount of impurity phases (i.e. $\alpha$-Fe grains, La-rich phases \cite*{katter2008magnetocaloric}$^,$ \cite*{bennati2015local}). The ingot, crushed in fragments with typical size smaller than 1 mm, is then full hydrogenated to raise the Curie point (about +150 K) and to ensure the long-term stability of the compound\cite*{Krautz2014}.

For the experiments we selected a single fragment of 5.26 mg with a flat surface. To check the compositional spread of the selected fragment over a mm scale, the elemental composition of the magnetocaloric phase was analysed by Energy Dispersive X-ray Spectroscopy (EDS) with a ZAF standardless quantification routine. The results of elemental semi-quantitative analysis are reported in Tab.\ref{tab1}. 
The standard deviation measured for Mn, which is the element present in lower concentration in the sample, and the most influential on the transition temperature, is 5$\%$ relative, which is compatible with the repeatability limit obtained on standard specimens of metallic alloys of known and uniform composition \cite{std}. This suggests that the actual compositional spread of the analysed sample (instrumental and counting statistics factors excluded) is negligible and the magnetocaloric phase of the studied fragment is expected to have a single transition temperature.
  
\begin{table}[h]
	\small
	\begin{tabular}{|c|c|c|c|c|}
		\hline
		\multicolumn{5}{|c|}{\textbf{LaFe$_{11.41}$Mn$_{0.30}$Si$_{1.29}$-H$_{1.65}$} } \\
		\hline
	at.$\%$	& \textbf{La}  &  \textbf{Fe} & \textbf{Mn} & \textbf{Si} \\
		\hline
		\textit{Nominal values}&	7.14 & 81.5	 &	2.14& 9.21\\
		\hline
		\textit{Measured values} 	&	7.40 & 79.60&	2.09& 10.92	\\
		\hline
		\textit{St.dev.} 	& 0.18 & 0.49 &	0.11 & 	0.49	\\
		\hline
	\end{tabular}	\caption{\label{tab1}: Atomic $\%$ composition of the main MCE phase of the investigated samples: nominal values are compared to the microanalysis values obtained on the selected fragment. Hydrogen atoms cannot be detected by the EDS technique. }
\end{table}
The in-temperature experiments were performed employing a differential scanning calorimeter based on two Peltier cells, a reference one and the sample holder, on which the specimen has been attached with silver paint \cite*{kuepferling2013rate}. The two cells are connected differentially in order to subtract the common heat flux background, thus the heat flux exchanged with the sample is computed from the voltage difference measured at the end of the two cells. 
The important quantities returned from the calorimeter are the heat flux, $q_s$, and the temperature of the sample holder, $T_p$. All the details about the setup can be found in reference \cite*{Basso2010}. Electrical measurements were performed, on the same sample, in a two stage cryogen-free cryocooler using a standard four point technique and monitoring the temperature of the cryostat by a Cernox{\tiny$^{\rm {TM}}$} thermometer. 

\medskip

\begin{figure}[h]
\centering
\includegraphics[width=0.85\linewidth]{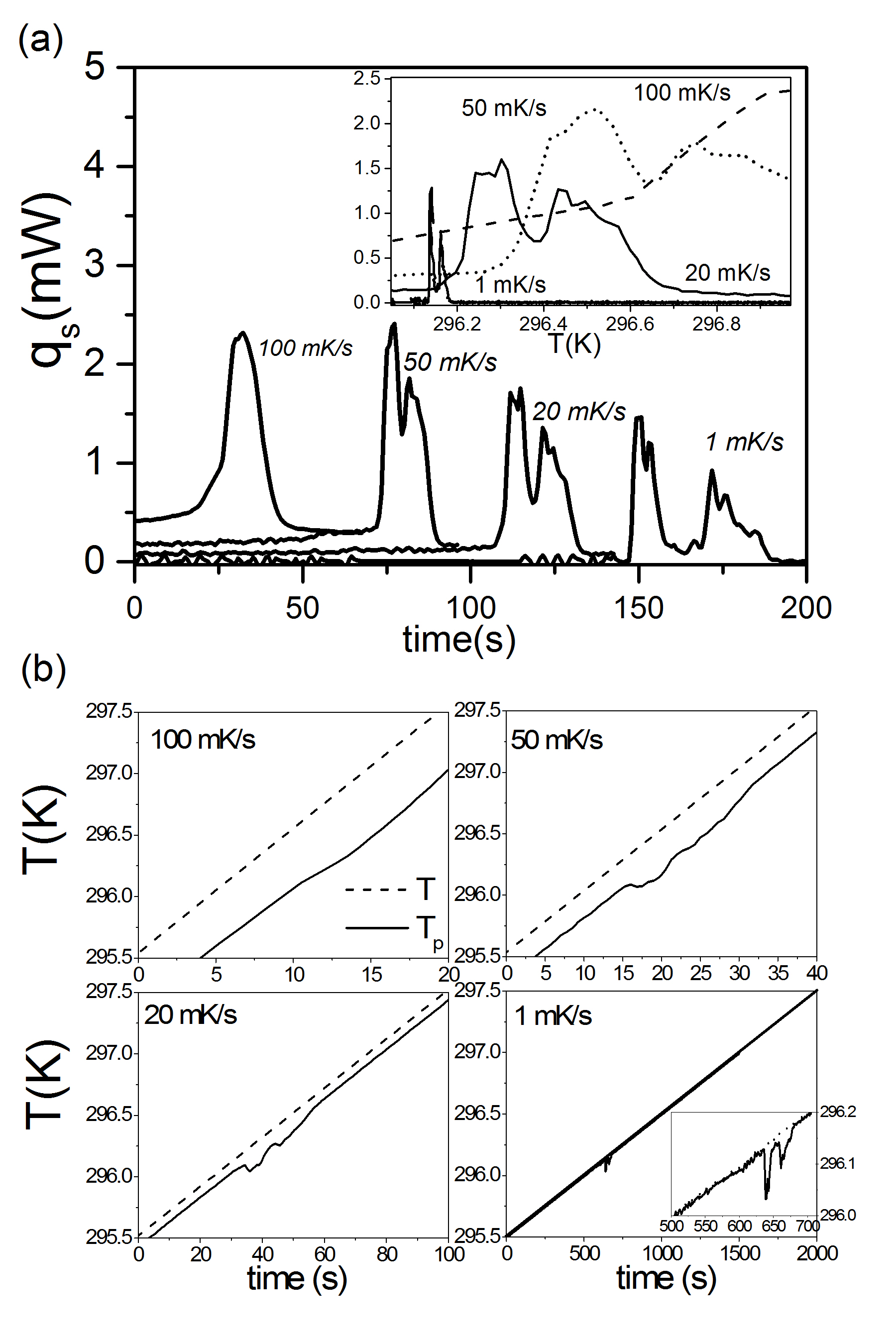}
\caption{(a) Temperature induced phase transitions at different scanning rates as a function of time. The magnetic field is fixed at $H$ = 0. Time values are translated to permit the complete visualization of all the transitions. Inset: same phase transitions as a function of the thermal bath temperature. (b) Temperature as a function of time for different scanning rates (100, 50, 20 and 1 mK/s) on heating: $T$ is the temperature of the thermal bath - monitored by a Pt100 thermometer-, whereas $T_p$ is the temperature of the sample holder.}
\label{fig:graph1}
\end{figure}

Zero magnetic field transitions obtained by calorimetric measurements while controlling and varying the thermal bath temperature, $T$, at different heating rates (d$T$/dt = 100 mK/s, 50 mK/s, 20 mK/s and 1 mK/s) are presented in the inset graph of Fig.\ref{fig:graph1} (a). The main figure shows that, slowing down d$T$/dt, the heat flux signal as a function of time becomes structured into a succession of individual transformation events. These isolated events are not visible in the faster ($>$ 50 mK/s) scanning rate, suggesting that they are related to sudden transformations of volume, which are likely associated to switches between metastable states.
Fig.\ref{fig:graph1} (b) shows the plots of $T$ and of the sample holder temperature, $T_p$, as a function of time for the same heating sequence of Fig.\ref{fig:graph1} (a). Away from the transition region, the lag between the two temperatures is constant depending only on the scanning rate and it becomes almost null for the 1 mK/s measurements. In the transition region, only $T_p$ has sudden drops, similarly to overheating processes. When an avalanche starts, and the sample begins to transform, $T_p$ is rapidly decreased of several mK, then the whole system tends to restore an equilibrium with the thermal bath temperature. 
The low rate measurements help to resolve a multiple peaks signal which may be due to different factors, for example, slight differences in the composition across the volume giving rise to a distribution of transition temperatures \cite*{morrison2010contributions}. Furthermore, the microstructural disorder (grain boundaries, magnetic and non magnetic precipitated phases \cite*{bennati2015local}) may locally modify the energy landscape facing the phase front, representing nucleation and/or pinning sites \cite*{kuepferling2014dynamics}$^,$\cite{imry1979influence}. 

\begin{figure}[h]
 	\centering
 	\includegraphics[width=0.85\linewidth]{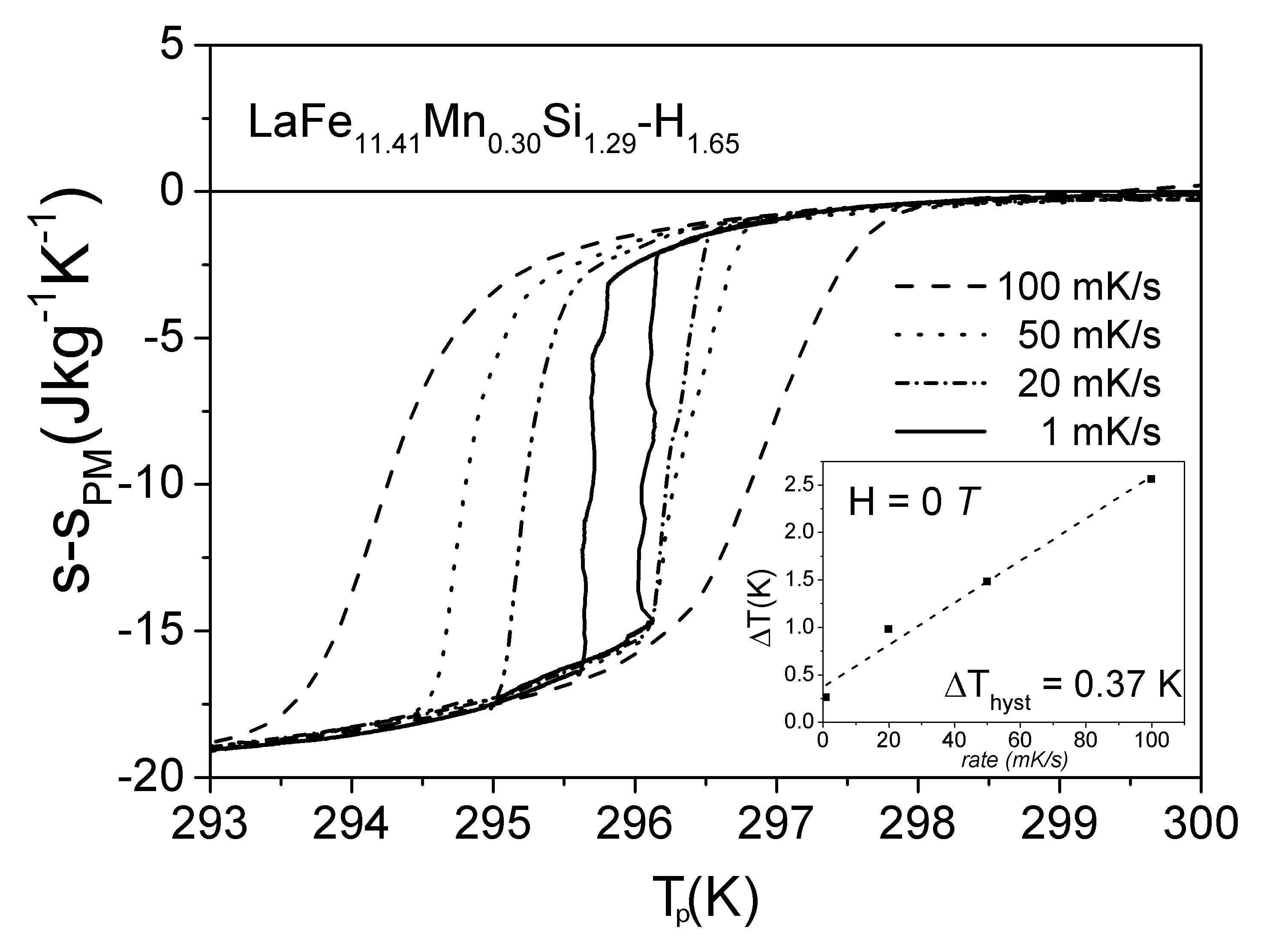}
 	\caption{Total entropy change of the transition at different applied scanning rates (100, 50, 20 and 1 mK/s). Inset: thermal hysteresis in 0 $H$ as a function of the temperature scanning rate. S$_{PM}$ is the entropy of the paramagnetic state.  }
 	\label{fig:graph2}
 \end{figure}

The effective entropy changes, computed for the four scan rates, are compared in Fig.\ref{fig:graph2}.
A linear fit of thermal hysteresis versus the scanning rate (inset graph at Fig.\ref{fig:graph2}) yields a $\textquotedblleft$ zero rate $\textquotedblright$ hysteresis of the compound of 0.37 K. This value is smaller but close to the 0.4 K obtained for agglomerated fragments (total mass = 50 mg) of the same material closed in an aluminium pan \cite*{basso2015specific}. Moreover, the $\Delta S$ value keeps nearly unchanged independently of the rate of scan, but at 1 mK/s, it is shown that the entropy follows a re-entrant hysteresis loop as a function of $T_p$. This effect reflects either the multi-avalanche heat flux signal of Fig.\ref{fig:graph1} (a) and the plot of the temperature vs time of Fig.\ref{fig:graph1} (b).
The low rate technique thus permits to quantify the important MCE quantities of the material and, in addition, to observe microscopic details of the phase transition that are potentially useful to determine the typical avalanche sizes and to understand their intrinsic kinetics \cite*{moore2009reducing}$^,$\cite*{lovell2015dynamics}.

\begin{figure}[h]
	\centering
	\includegraphics[width=0.85\linewidth]{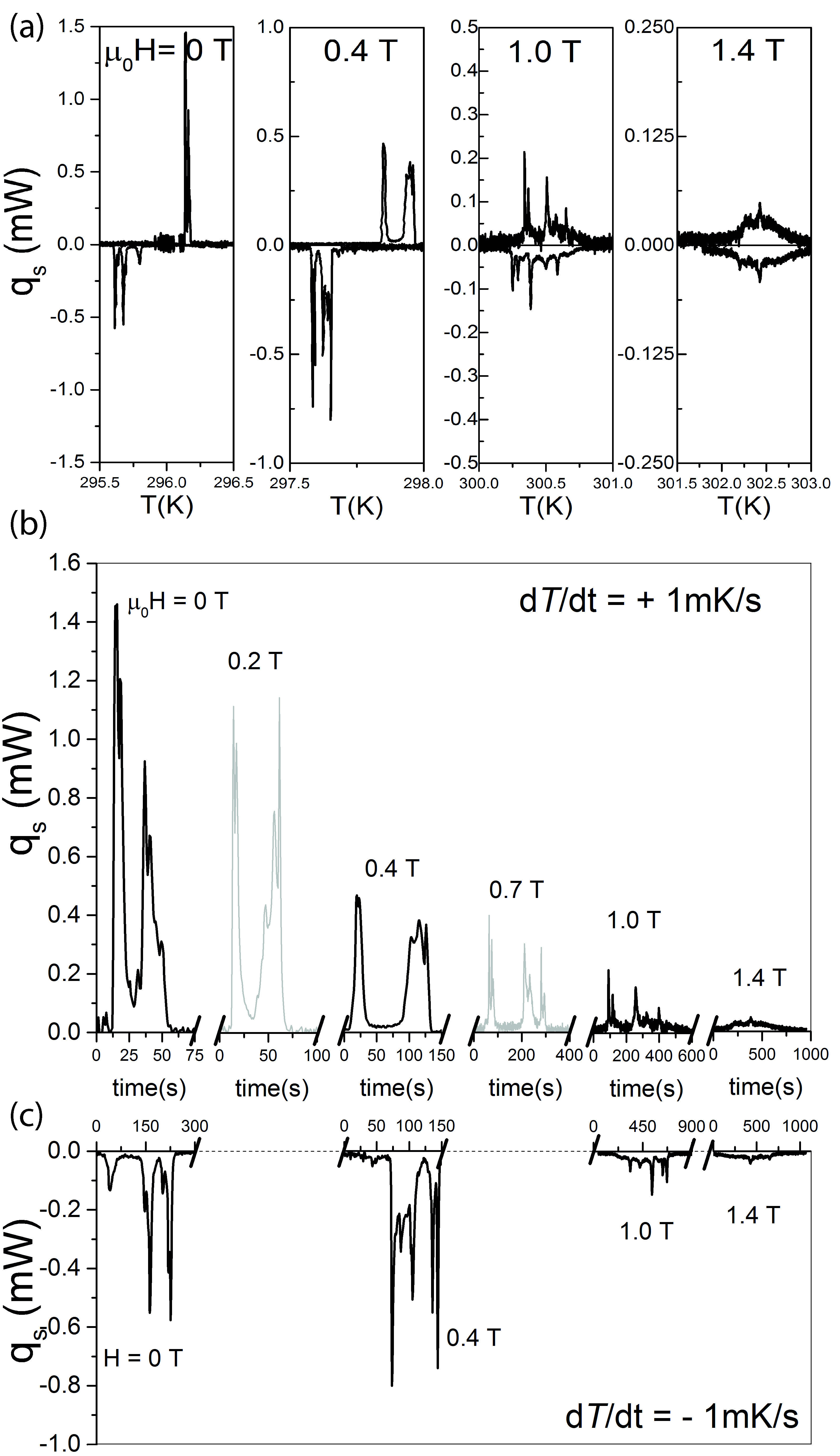}
	\caption{(a) Temperature induced phase transitions at different applied magnetic fields as a function of the temperature of the thermal bath, $T$; the scanning rate is 1 mK/s. (b) and (c) heating and cooling, respectively, temperature-induced phase transitions at different applied magnetic fields as a function of the time. Time has been translated to permit the complete visualization of the transitions.}
	\label{fig:graph3}
\end{figure}

In order to deepen all the features observed in the low rate scans of the zero field transition, the evolution of the avalanches as function of temperature was studied by applying different magnetic fields, being able to cross the critical point for which the transition becomes of the second order \cite*{basso2015specific}. The top graph on Fig.\ref{fig:graph3} (a) displays the transitions as a function of $T$ on heating (FM to PM, endothermic) and on cooling (PM to FM, exothermic) obtained in magnetic fields of 0, 0.4, 1.0 and 1.4 tesla. 
Two main features appear with field: i) the avalanches change in number and in magnitude, ii) a continuous background gradually substitutes the heat flux content of the peaks. 
The heating sequence plotted as a function of time in Fig.\ref{fig:graph3} (b) shows how the heat flux avalanches are gradually suppressed by the magnetic field. The cooling sequence of Fig.\ref{fig:graph3} (c) is analogous to the heating one and, correspondingly, it shows a redistribution and a gradual reduction of the heat flux avalanches with increasing field. The falling magnitude of the peaks reflects both the lowering of the latent heat content of the transition \cite*{morrison2010contributions} and the gradual reduction of the overall free energy barrier between the FM and the PM phase \cite*{fujita2001itinerant} with field. Beside their magnitude values, we observe that the avalanches of the higher fields measurements, do not appear as just shifted in temperature with respect the zero-field avalanches. According to the EDS results we can thus exclude a significant graded distribution of $T_t$ due to non uniform chemical compositions. Instead, we suppose a nucleation mechanism which changes relevantly near the critical point ($H_c$,$T_c$).

We interpret the varying number of avalanches as follows: grain boundaries, precipitated phase or other sources of disorder are preferential nucleation sites and are randomly distributed into the sample; if the energy of the system, driven by temperature, overwhelms the local energy barrier between phases at these sites, an avalanche can be set off. By reducing the overall intrinsic energy barrier with field, a new modulation of the local energies values is expected and the avalanches in our experiments change: they increase in number and decrease in amplitude. The local energy barrier to overcome indeed depends on both the intrinsic PM/FM energy barrier (characteristic of the compound) and on the local disorder (characteristic of the sample). In such a picture, approaching the critical point, the decreased free-energy barrier takes advantage of local fluctuations \cite*{imry1979influence}.

\begin{figure}[h]
\centering
\includegraphics[width=0.85\linewidth]{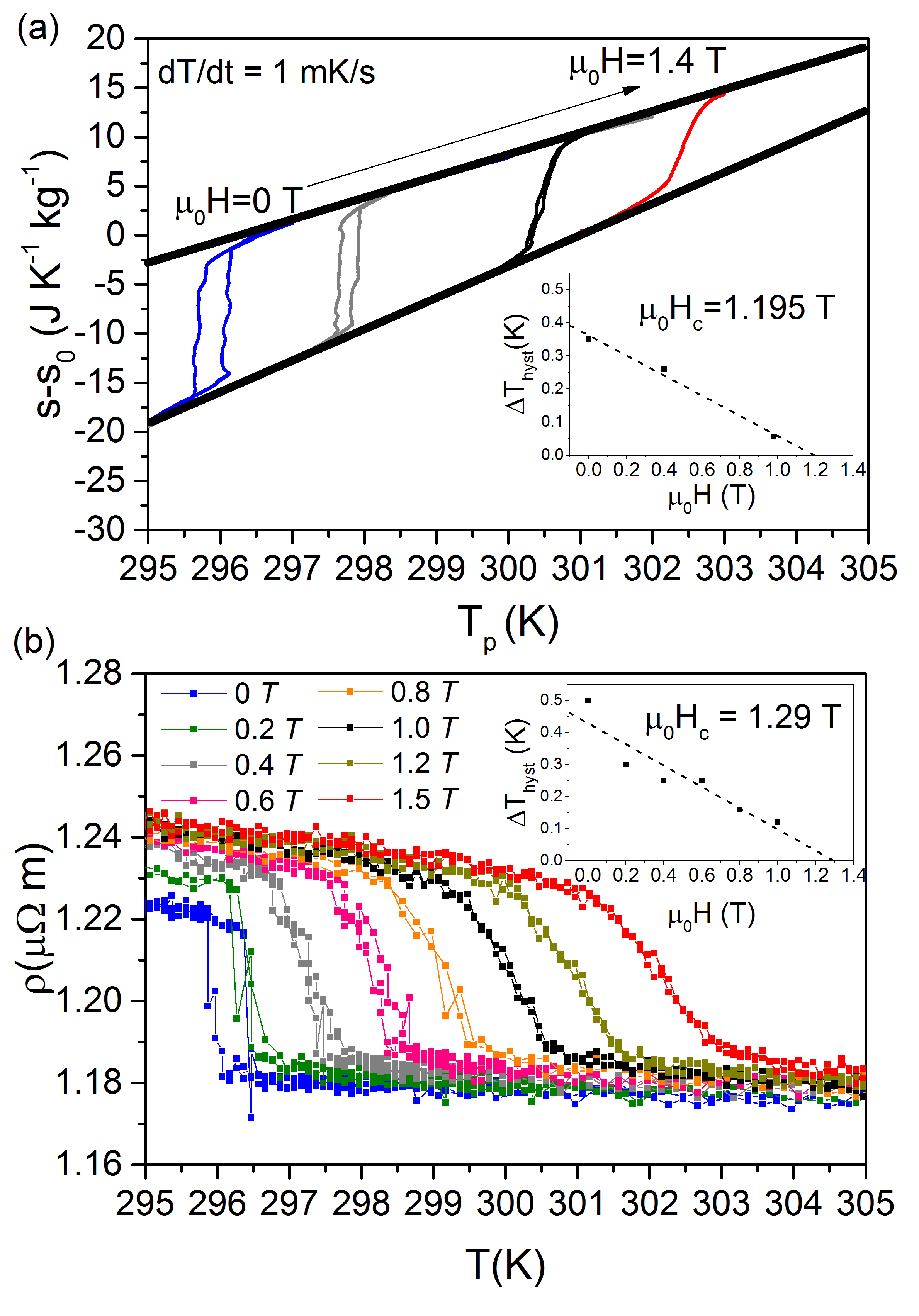}
\caption{Total entropy variation (a) and electrical resistivity behaviour (b) at the temperature induced phase transitions in different applied magnetic field as a function of temperature; scanning rates are respectively 1 mK/s and $\sim$ 3.5 mK/s. Insets: the relative thermal hysteresis as a function of magnetic field.}
\label{fig:graph4}
\end{figure}

For completeness, the temperature dependence of the electrical resistivity, $\rho$, at different applied magnetic fields was measured and the results are plotted in Fig.\ref{fig:graph4} (b). The FM and PM phases have different values of $\rho$, mainly ascribable to the larger cubic cell of the low temperature phase \cite*{palstra1983}$^,$\cite*{hu2004}, thus that any transformed fraction of the volume can be thought as a new resistive element in the measurements. It is worth to observe that transitions below the critical magnetic field of 1.29 $\pm$ 0.17 T show detectable jumps in $\rho$, which are again a sign of the heterogeneous transformation of volumes inside the sample. 

Temperature hysteresis values had been computed from entropy changes for $\mu_0 H \neq$ 0 (Fig.\ref{fig:graph4} (a)) providing a critical field value of about $\mu_0 H_c$ = 1.195 T (inset graph). This value and the one obtained from faster rates measurements ($H_c$ = 1.3 T \cite*{basso2015specific}) agree within the experimental errors with that obtained from electrical resistivity (the critical magnetic field is about 1.29 $\pm$ 0.17 T). The most precise value of the zero field $\Delta T_{hyst}$= 0.37 K has been achieved with the low rate calorimetric technique.

We have thus shown, with different experimental detection techniques, that the evolution of the first order transitions in our La(Fe,Si)$_{13}$ based material is characterized by burst-like events separated by inactivity periods. The transition appears very different depending whether it is close to or far from the critical point. A temperature scan rate as low as 1 mK/s is able to highlight the interplay between intrinsic energy barrier and structural disorder.  This finding opens up further questions on the kinetics of the phase transition, disclosing the existence of a rough and complex energy landscape which is critical in determining the onset of transitions in La(Fe,Si)$_{13}$ based compounds. Theoretical models aimed to describe the kinetics of the avalanches and its relevance to the dynamic behaviour of the magnetocaloric effect are underway. These models should take into account the fact that the free energy of the system depends on both the intrinsic barrier as well as on extrinsic contributions due to local perturbations induced by quenched disorder.

\medskip

We thank Vacuumschmelze GmbH $\&$ Co for the samples.

\bibliography{bibcopia}

\end{document}